\documentclass[12pt,a4paper,superscriptaddress]{revtex4}
\usepackage{graphicx}
\usepackage{graphics}
\usepackage{amsmath}
\usepackage{dcolumn}
\usepackage{amssymb}
\usepackage{bm}
\usepackage[latin1]{inputenc}
\usepackage[colorlinks,bookmarks=true]{hyperref}
\newcommand{\be}{\begin{equation}}
\newcommand{\ee}{\end{equation}}
\newcommand{\bea}{\begin{eqnarray}}
\newcommand{\eea}{\end{eqnarray}}
\newcommand{\hf}{\frac{1}{2}}

\begin{document}

\title{On the one-loop effective potential in the higher-derivative four-dimensional chiral superfield theory with a nonconventional kinetic term}

\author{F. S. Gama}
\email{fgama@fisica.ufpb.br}
\affiliation{Departamento de F\'{\i}sica, Universidade Federal da Para\'{\i}ba\\
 Caixa Postal 5008, 58051-970, Jo\~ao Pessoa, Para\'{\i}ba, Brazil}

\author{M. Gomes}
\email{mgomes@fma.if.usp.br}
\affiliation{Departamento de F\'{\i}sica Matem\'atica, Universidade de S\~ao Paulo,\\
 Caixa Postal 66318, 05314-970, S\~ao Paulo, SP, Brazil}

\author{J. R. Nascimento}
\email{jroberto@fisica.ufpb.br}
\affiliation{Departamento de F\'{\i}sica, Universidade Federal da Para\'{\i}ba\\
 Caixa Postal 5008, 58051-970, Jo\~ao Pessoa, Para\'{\i}ba, Brazil}

\author{A. Yu. Petrov}
\email{petrov@fisica.ufpb.br}
\affiliation{Departamento de F\'{\i}sica, Universidade Federal da Para\'{\i}ba\\
 Caixa Postal 5008, 58051-970, Jo\~ao Pessoa, Para\'{\i}ba, Brazil}
 
\author{A. J. da Silva}
\email{ajsilva@fma.if.usp.br}
\affiliation{Departamento de F\'{\i}sica Matem\'atica, Universidade de S\~ao Paulo,\\
 Caixa Postal 66318, 05314-970, S\~ao Paulo, SP, Brazil}

\begin{abstract}
We explicitly calculate the one-loop effective potential for a higher-derivative four-dimensional chiral superfield theory with a nonconventional kinetic term. We consider the cases of minimal and nonminimal general Lagrangians. In particular, we find that in the minimal case the divergent part of the one-loop effective potential vanishes by reason of the chirality.
\end{abstract}

\maketitle

\section{Introduction}
Higher-derivative field theories have aroused  great interest by many reasons. The main of them is certainly the fact that the presence of higher derivatives improves the renormalization behaviour. This 
 called the attention to these theories, in the gravity context where  higher-derivative additive term were used to construct  renormalizable theories of  gravity \cite{Stelle}. Further studies of gravity with additive higher-derivative terms, which naturally emerge within the anomaly context \cite{AM}, led to the possibility of supersymmetric extensions. The first example of such theories is the dilaton supergravity \cite{dilaton} based on the results for the supertrace anomaly \cite{anomaly}. In its essence, the dilaton supergravity, in which the dynamics is completely described by only nontrivial supergravity prepotential, so-called chiral compensator \cite{SGRS}, was the first higher-derivative superfield theory discussed in details at the quantum level.

Further, the equivalence of higher-derivative (super)field theories to theories without higher derivatives but with additional ghost fields has been shown at the tree level for several types of  models \cite{Ant}. Therefore, the interest to the general properties of the effective action of higher-derivative superfield theories arose. Another important reason of the interest in higher-derivative field theories is the concept of the Horava-Lifshitz gravity \cite{HL} which  involves higher derivatives, although only in the spatial sector, and without ghosts.

In the papers \cite{ourhigh} a higher-derivative theory with a conventional kinetic term (in which the higher derivatives enter the general Lagrangian in the form $\int d^8 z\Phi \Box\bar{\Phi}$) has been studied, the one-loop low-energy effective action has been considered, and its super-renormalizability (and, in certain cases, finiteness) has been shown. Further, similar studies have been done for the four-dimensional supergauge theory \cite{Gama} and for the three-dimensional superfield theories \cite{FSG3D}. 

In this paper, we extend these studies by obtaining the one-loop effective potential for the chiral superfield theory with a nonconventional kinetic term. It is characterized by the presence of higher derivatives in a classical chiral Lagrangian. Up to now, such a theory was treated only in the paper \cite{Ant} where it was studied on the tree level in the context of equivalence with the theories without higher derivatives but with additional ghost fields. Its quantum properties never have been studied. This is just the problem we address in this paper.

To do this study, we use the methodology of obtaining the superfield effective potential originally developed in \cite{BKY}. Furthermore this method has been successfully applied in
\cite{EP,ourhigh,Gama} for different superfield models. Throughout this paper, we use notations and conventions from the book \cite{BK0}.

\section{Minimal general Lagrangian}
Let us start with the following $4D$ higher-derivative superfield theory where the higher derivatives are implemented in a nonconventional way, that is, in the chiral Lagrangian:
\bea
\label{nm}
S=\int d^8z \Phi\bar{\Phi}+[\int d^6z(-\frac{a}{2}\Phi\Box\Phi+W(\Phi))+h.c.].
\eea
Here $a$ is a constant with a negative $(-1)$ mass dimension, {and the chiral potential is given by an arbitrary holomorphic function $W(\Phi)$ which in some cases will be particularized as $W(\Phi)=\hf m\Phi^2+\frac{\lambda}{3!}\Phi^3$. We remind that earlier, theories of such a form have been studied only at the tree level \cite{Ant}, whereas the quantum calculations were carried out only for higher derivatives in the general effective Lagrangian, that is, just the case discussed in details in \cite{ourhigh}.

 Before  performing any quantum calculation, it is convenient to determine the superficial degree of divergence $\omega$ for the model (\ref{nm}). First, we have to calculate the propagators for (\ref{nm}). They are given by
\bea
\langle\bar\Phi(1)\Phi(2)\rangle&=&\frac{i}{k^2+(m+a k^2)^2}\delta_{12} \ ,\\
\langle\Phi(1)\Phi(2)\rangle&=&i\frac{(m+a k^2)D^2_1}{4k^2[k^2+(m+a k^2)^2]}\delta_{12} \ ,\\
\langle\bar\Phi(1)\bar\Phi(2)\rangle&=&i\frac{(m+a k^2)\bar D^2_1}{4k^2[k^2+(m+a k^2)^2]}\delta_{12} \ ,
\eea
where $m=W^{\prime\prime}(\Phi)|_{\Phi=0}$.
Second, we have to determine the  number of covariant derivatives in each of the supergraphs. In our case, the number is the same as that of the Wess-Zumino model, namely $4V-2E+2C-4L$, where $V$, $E$, $P$, $C$, and $L$ denote the number of vertices, external lines, $\langle\bar\Phi(1)\Phi(2)\rangle$-propagators, chiral propagators, and loops, respectively.

Finally, we assume that all covariant derivatives are converted into momenta via the $D$-algebra. Then, it follows that the maximal superficial degree of divergence is given by
\bea
\omega_{max}&=&4L-4P-4C+\hf(4V-2E+2C-4L) \nonumber\\
&=&2-2P-C-E \ ,
\eea
where we used the topological identity $L+V-P-C=1$. Of course, for any non-trivial supergraph contribution to the effective action we will have $E\ge2$ and $P$ or $C$ are no less than 1. Therefore, the theory (\ref{nm}) is finite which is natural for  higher-derivative models (cf. \cite{dilaton}).

Now, let us begin with the calculation of the one-loop effective potential. In order to find the one-loop effective action we use a trick originally introduced in \cite{BKY}to transform  the integral over chiral superfields to the integral over unconstrained superfields, that is, we introduce the effective action $W$ for the free gauge superfield $v$:
\bea
\label{w}
e^{iW}=\int Dv\,\delta(\frac{1}{4}D^2v-\bar{\phi})\delta(\frac{1}{4}\bar{D}^2v-\phi)\exp(-\frac{i}{16}\int d^8z v D^{\alpha}\bar{D}^2D_{\alpha}v).
\eea
It is clear that $W$ is a constant (indeed, it is gauge independent, so, it does not depend on the gauge fixing functions, and, thus, on $\phi$ and $\bar{\phi}$). Then, let us do a background-quantum splitting in the action (\ref{nm}), by the rule $\Phi\to\Phi+\phi$, $\bar{\Phi}\to\bar{\Phi}+\bar{\phi}$, where $\Phi,\bar{\Phi}$ are from  now on the background chiral and antichiral fields. The quadratic part in the  quantum fields  of the action (\ref{nm}) is
\bea
S_2=\int d^8z \phi\bar{\phi}+[\frac{1}{2}\int d^6z(-a\phi\Box\phi+\Psi\phi^2)+h.c.],
\eea
where $\Psi=W^{\prime\prime}(\Phi)$ is a background chiral field.
The one-loop effective action corresponding to the theory (\ref{nm}) is $\Gamma^{(1)}$ defined by
\bea
e^{i\Gamma^{(1)}[\Phi,\bar{\Phi}]}&=&\int D\phi D\bar{\phi}\exp (iS_2)=\int D\phi D\bar{\phi}\exp (i\Big[\int d^8z \phi\bar{\phi}+\nonumber\\&+&
[\frac{1}{2}\int d^6z(-a\phi\Box\phi+\Psi\phi^2)+h.c.]\Big]).
\eea
Let us multiply this expression with the equation (\ref{w}). Using functional delta functions and disregarding irrelevant constants, we get
\bea
e^{i\Gamma^{(1)}[\Phi,\bar{\Phi}]}&=&\int D\phi D\bar{\phi}\exp (iS_2)=\int Dv\exp \Big[i\frac{1}{2}v\int d^8z\big(\Box -
\nonumber\\&-& \frac{a}{4}\Box(D^2+\bar{D}^2)+\frac{1}{4}(\Psi\bar{D}^2+\bar{\Psi} D^2)\big)v\Big].
\eea
So, we find that the desired one-loop effective action is
\bea
\Gamma^{(1)}=-\frac{i}{2}{\rm Tr}\ln\big[\Box+\frac{1}{4}(-a\Box+\Psi)\bar{D}^2+\frac{1}{4}(-a\Box+\bar{\Psi})D^2\big].
\eea
So, it remains to calculate this trace of the logarithm. To do it, we expand it in power series:
\bea
\Gamma^{(1)}=-\frac{i}{2}{\rm Tr}\sum_{n=1}^{\infty}\frac{(-1)^{n+1}}{n}\Big[\frac{(-a\Box+\Psi)\bar{D}^2+(-a\Box+\bar{\Psi})D^2}{4\Box}\Big]^n.
\eea
It is easy to see that, for  $A$ and $\bar{A}$ constant and arbitrary (this is just the case sufficient for the calculation of the superfield effective potential where the background superfields are constant \cite{BKY}),
\bea
{\rm Tr}\big[\frac{1}{4}(\bar{A}D^2+A\bar{D}^2)\big]^{2k}=\int d^{8}z\,(A\bar{A}\Box)^k\frac{2}{\Box}\delta^4(x-x')|_{x=x'}.
\eea
For an  odd number of multipliers the trace is zero. Thus,
\bea
\Gamma^{(1)}=\frac{i}{2}\int d^8z\sum_{k=0}^{\infty}\frac{1}{k}\Big[\frac{(-a\Box+\Psi)(-a\Box+\bar{\Psi})}{\Box}\Big]^k
\frac{1}{\Box}\delta^4(x-x')|_{x=x'},
\eea
after Fourier transform, Wick rotation and summation, {we arrive at the following contribution to the K\"{a}hlerian
effective potential
\bea
\label{k1}
K^{(1)}(\bar\Phi,\Phi)=-\frac{1}{2}\int\frac{d^4p}{(2\pi)^4}\frac{1}{p^2}
\ln\Big[1+\frac{(a p^2+\Psi)(a p^2+\bar{\Psi})}{p^2}\Big].
\eea
It is clear that at $a=0$, we restore the well-known expression $K^{(1)}(\bar\Phi,\Phi)=-\frac{1}{2}\int\frac{d^4p}{(2\pi)^4}\frac{1}{p^2}
\ln(1+\frac{\Psi\bar{\Psi}}{p^2})$ (it was explicitly found in \cite{BKY} where it was shown that in this case, after subtraction of the UV divergences, one has $K^{(1)}(\bar\Phi,\Phi)=-\frac{1}{32\pi^2}\Psi\bar{\Psi}\ln\frac{\Psi\bar{\Psi}}{\mu^2}$). So, it remains to do the integral in (\ref{k1}). This task can be done by splitting the logarithm in three parts,
\bea
\label{int}
K^{(1)}(\bar\Phi,\Phi)=-\hf \int \frac{d^4p}{(2\pi)^4}\frac{1}{p^2}\Big[\ln{(p^2+\Omega_+)}+\ln{(p^2+\Omega_-)}-\ln{(p^2)}\Big] \ ,
\eea
where
\bea
\label{omega}
\Omega_{\pm}=\frac{1+a(\bar\Psi+\Psi)\pm\sqrt{[1+a(\bar\Psi+\Psi)]^2-4a^2\bar\Psi\Psi}}{2a^2} \ .
\eea
The last logarithm does not depend on the background superfields, then we can drop it  by means of the normalization of the effective action. The integrals in (\ref{int}) are well-known and can be computed by using the dimensional reduction prescription. Therefore, we regularize this integral by the formal replacement of $d^4p$ by $\mu^{4-2\omega}d^{2\omega}p$, so, one has 
\bea
\label{int1}
K^{(1)}(\bar\Phi,\Phi)=-\hf \mu^{4-2\omega}\int \frac{d^{2\omega}p}{(2\pi)^{2\omega}}\frac{1}{p^2}\Big[\ln{(p^2+\Omega_+)}+\ln{(p^2+\Omega_-)}\Big] \ ,
\eea
In the limit $\omega\rightarrow2$ we find
\bea
K^{(1)}(\bar\Phi,\Phi)=\frac{1}{32\pi^2(2-\omega)}(\Omega_++\Omega_-)-\frac{1}{32\pi^2}\Big[\Omega_+\ln{\Big(\frac{\Omega_+}{\mu^2}\Big)}+\Omega_-\ln{\Big(\frac{\Omega_-}{\mu^2}\Big)}\Big] \ ,
\eea
where dimensionless constants were removed by means of a redefinition of the parameter $\mu^2$.

Although the first term is divergent, we notice from (\ref{omega}) that $(\Omega_++\Omega_-)=\frac{1+a(\bar\Psi+\Psi)}{a^2}$. Due to the fact that the contribution to the effective action is obtained by the integration of the K\"{a}hlerian potential over $d^8z$, we find that the divergent term is annihilated by the Grassmann integration. Therefore, the final result is finite and equal to
\bea
\label{minimalfinalresult}
K^{(1)}(\bar\Phi,\Phi)&=&-\frac{1}{32\pi^2}\Big[\frac{1+a(\bar W^{\prime\prime}+W^{\prime\prime})+\sqrt{[1+a(\bar W^{\prime\prime}+W^{\prime\prime})]^2-4a^2\bar W^{\prime\prime}W^{\prime\prime}}}{2a^2}\nonumber\\
&\times&\ln{\Big(\frac{1+a(\bar W^{\prime\prime}+W^{\prime\prime})+\sqrt{[1+a(\bar W^{\prime\prime}+W^{\prime\prime})]^2-4a^2\bar W^{\prime\prime}W^{\prime\prime}}}{2\mu^2a^2}\Big)}\nonumber\\
&+&\frac{1+a(\bar W^{\prime\prime}+W^{\prime\prime})-\sqrt{[1+a(\bar W^{\prime\prime}+W^{\prime\prime})]^2-4a^2\bar W^{\prime\prime}W^{\prime\prime}}}{2a^2}\nonumber\\
&\times&\ln{\Big(\frac{1+a(\bar W^{\prime\prime}+W^{\prime\prime})-\sqrt{[1+a(\bar W^{\prime\prime}+W^{\prime\prime})]^2-4a^2\bar W^{\prime\prime}W^{\prime\prime}}}{2\mu^2a^2}\Big)}\Big] \ .
\eea
It should be noticed that, in spite of the explicit dependence of this expression on $\mu$, the corresponding low energy effective action is scale invariant.

To check the consistency of our result, we take the limit $a\rightarrow0$ in order to get the K\"{a}hlerian
potential for the Wess-Zumino model. Hence, from (\ref{minimalfinalresult}) we find
\bea
K^{(1)}(\bar\Phi,\Phi)&=&-\frac{1}{32\pi^2}\Bigg\{\frac{1}{a^2}\ln{\Big(\frac{1}{\mu^2a^2}\Big)}+\frac{1}{a}\bigg[\bar W^{\prime\prime}\ln{\Big(\frac{e}{\mu^2a^2}\Big)}
+W^{\prime\prime}\ln{\Big(\frac{e}{\mu^2a^2}\Big)+\frac{a}{2}\big((\bar W^{\prime\prime})^2}\nonumber\\
&&+(W^{\prime\prime})^2\big)\bigg]-\bar W^{\prime\prime}W^{\prime\prime}\ln{\Big(\frac{1}{\mu^2a^2}\Big)+\bar W^{\prime\prime}W^{\prime\prime}\ln{\Big(\frac{\bar W^{\prime\prime}W^{\prime\prime}}{\mu^2}\Big)}}+\mathcal{O}(a)\Bigg\} \ .
\eea
Again, due to the fact that the K\"{a}hlerian potential must be integrated over $d^8z$, we conclude that  the only nontrivial terms are
\bea
K^{(1)}(\bar\Phi,\Phi)&=&-\frac{1}{32\pi^2}\bar W^{\prime\prime}W^{\prime\prime}\ln{(a^2\bar W^{\prime\prime}W^{\prime\prime})}\ .
\eea
We notice that this is the one-loop effective potential for Wess-Zumino model, as expected, with its characteristic divergence,  when $a\rightarrow0$.

\section{Nonminimal general Lagrangian}

Now, let us consider the more generic case, that is, the arbitrary general Lagrangian $K(\bar\Phi,\Phi)$ (we will refer to this case as to the nonmimimal one, as it was called in \cite{EP}):
\bea
S=\int d^8zK(\bar\Phi,\Phi)+\bigg[\int d^6z(-\frac{a}{2}\Phi\Box\Phi+W(\Phi))+h.c.\bigg] \ .
\eea
Repeating the calculations from the previous section, we find that the effective potential for this model is given by:
\bea
\label{the2}
K^{(1)}(\bar\Phi,\Phi)=-\int \frac{d^4p}{(2\pi)^4}\frac{1}{p^2}\ln{(K_{\bar\Phi\Phi})}-\hf \int \frac{d^4p}{(2\pi)^4}\frac{1}{p^2}\ln{\Big[1+\frac{(a p^2+\bar W^{\prime\prime})(a p^2+W^{\prime\prime})}{K_{\bar\Phi\Phi}^2p^2}\Big]} \ .
\eea
The first term in this expression vanishes within the dimensional reduction scheme. The second one can be simplified by use of the identity
\bea
1+\frac{(\widetilde{a} p^2+\widetilde{\bar W^{\prime\prime}})(\widetilde{a}p^2+\widetilde{W^{\prime\prime}})}{p^2}=\frac{(p^2+\Omega_+)(p^2+\Omega_-)}{p^2}\tilde{a}^2 \ ,
\eea
where we denoted $\widetilde{Q}\equiv \frac{Q}{K_{\bar\Phi\Phi}}$ (with $Q=W^{\prime\prime}$ or $Q=\bar{W}^{\prime\prime}$ or $Q=a$), and
\bea
\Omega_{\pm}=\frac{1+\widetilde{a}(\widetilde{\bar W^{\prime\prime}}+\widetilde{W^{\prime\prime}})\pm\sqrt{[1+\widetilde{a}(\widetilde{\bar W^{\prime\prime}}+\widetilde{W^{\prime\prime}})]^2-4\widetilde{a}^2
\widetilde{\bar W^{\prime\prime}}\widetilde{W^{\prime\prime}}}}{2\widetilde{a}^2} \ .
\eea
Thus, we can rewrite (\ref{the2}) as
\bea
K^{(1)}(\bar\Phi,\Phi)=-\hf \int \frac{d^4p}{(2\pi)^4}\frac{1}{p^2}\Big[\ln{(p^2+\Omega_+)}+\ln{(p^2+\Omega_-)}-\ln{(p^2)}\Big] \ .
\eea
As before, the last term in this expression can be disregarded as it is  field independent. The other  integrals can be calculated by use of the dimensional reduction which again consists in the change  $d^4p\to\mu^{4-2\omega}d^{2\omega}p$. So, we have for $\omega\rightarrow2$
\bea
K^{(1)}(\bar\Phi,\Phi)=\frac{1}{32\pi^2(2-\omega)}(\Omega_++\Omega_-)-\frac{1}{32\pi^2}\Big[\Omega_+\ln{\Big(\frac{\Omega_+}{\mu^2}\Big)}+\Omega_-\ln{\Big(\frac{\Omega_-}{\mu^2}\Big)}\Big] \ ,
\eea
where some constants are reabsorbed into redefinition of $\mu^2$.

Finally, under appropriate substitutions, we find
\bea
\label{finalresult1}
K^{(1)}(\bar\Phi,\Phi)&=&\frac{K_{\bar\Phi\Phi}^2}{32\pi^2a^2(2-\omega)}+\frac{(\bar W^{\prime\prime}+W^{\prime\prime})}{32\pi^2a(2-\omega)}\nonumber\\
&-&\frac{1}{32\pi^2}\Big[\frac{K_{\bar\Phi\Phi}^2+a(\bar W^{\prime\prime}+W^{\prime\prime})+\sqrt{[K_{\bar\Phi\Phi}^2+a(\bar W^{\prime\prime}+W^{\prime\prime})]^2-4a^2\bar W^{\prime\prime}W^{\prime\prime}}}{2a^2}\nonumber\\
&\times&\ln{\Big(\frac{K_{\bar\Phi\Phi}^2+a(\bar W^{\prime\prime}+W^{\prime\prime})+\sqrt{[K_{\bar\Phi\Phi}^2+a(\bar W^{\prime\prime}+W^{\prime\prime})]^2-4a^2\bar W^{\prime\prime}W^{\prime\prime}}}{2\mu^2a^2}\Big)}\nonumber\\
&+&\frac{K_{\bar\Phi\Phi}^2+a(\bar W^{\prime\prime}+W^{\prime\prime})-\sqrt{[K_{\bar\Phi\Phi}^2+a(\bar W^{\prime\prime}+W^{\prime\prime})]^2-4a^2\bar W^{\prime\prime}W^{\prime\prime}}}{2a^2}\nonumber\\
&\times&\ln{\Big(\frac{K_{\bar\Phi\Phi}^2+a(\bar W^{\prime\prime}+W^{\prime\prime})-\sqrt{[K_{\bar\Phi\Phi}^2+a(\bar W^{\prime\prime}+W^{\prime\prime})]^2-4a^2\bar W^{\prime\prime}W^{\prime\prime}}}{2\mu^2a^2}\Big)}\Big] \ .
\eea
This result is explicitly divergent. However, we note that again, to obtain the contribution to the effective action, one must integrate the K\"{a}hlerian effective potential over $d^8z$. Thus, the second divergent term is annihilated via integration over Grassmannian coordinates, and the first one in the minimal case $K(\bar\Phi,\Phi)=\bar\Phi\Phi$ reduces to a field-independent constant whose integral over the superspace vanishes. At the same time, there is no other cases when the divergences vanish. Therefore, only the minimal models with $K(\bar\Phi,\Phi)=\bar\Phi\Phi$ yield the finite one-loop K\"{a}hlerian effective potential.

Taking the limit $a\rightarrow0$ in (\ref{finalresult1}), we find
\bea
K^{(1)}(\bar\Phi,\Phi)&=&-\frac{1}{32\pi^2}\Bigg\{\frac{1}{a^2}\bigg[-\frac{K_{\bar\Phi\Phi}^2}{(2-\omega)}+K_{\bar\Phi\Phi}^2\ln{\Big(\frac{K_{\bar\Phi\Phi}^2}{\mu^2a^2}\Big)}\bigg]
+\frac{1}{a}\bigg[W^{\prime\prime}\ln{\Big(\frac{eK_{\bar\Phi\Phi}^2}{\mu^2a^2}\Big)}\nonumber\\
&&+\bar W^{\prime\prime}\ln{\Big(\frac{e K_{\bar\Phi\Phi}^2}{\mu^2a^2}\Big)}\bigg]+\frac{1}{2K_{\bar\Phi\Phi}^2}(\bar W^{\prime\prime 2}+W^{\prime\prime 2})-\frac{\bar W^{\prime\prime}W^{\prime\prime}}{K_{\bar\Phi\Phi}^2}\ln{\Big(\frac{K_{\bar\Phi\Phi}^2}{\mu^2a^2}\Big)}\nonumber\\
&&+\frac{\bar W^{\prime\prime}W^{\prime\prime}}{K_{\bar\Phi\Phi}^2}\ln{\Big(\frac{\bar W^{\prime\prime}W^{\prime\prime}}{\mu^2K_{\bar\Phi\Phi}^2}\Big)}+\mathcal{O}(a)\Bigg\} \ .
\eea
For the usual case, when $K(\bar\Phi,\Phi)=\bar\Phi\Phi$ and $W(\Phi)=\frac{m}{2}\Phi^2+\frac{\lambda}{3!}\Phi^3$, the only terms surviving in the limit $a\rightarrow0$ (when the higher derivatives are "switched off") are
\bea
K^{(1)}(\bar\Phi,\Phi)&=&-\frac{1}{32\pi^2}\bar\Psi\Psi\ln{(a^2\bar\Psi\Psi)} \ ,
\eea
where $\Psi\equiv m+\lambda\Phi$.

Finally, notice that for $a\to 0$ we reproduced the one-loop effective potential for the Wess-Zumino model.

\section{Summary}

We calculated the one-loop effective potential for nonconventional higher-derivative chiral superfield models with minimal and nonminimal kinetic terms. We find that in the minimal case it displays a nonpolynomial, logarithmic-like behaviour while its finiteness is achieved in a very nontrivial manner -- the divergent part vanishes due to properties of the integral over the Grassmann variables (in the nonminimal case the finiteness does not occur, actually the theory becomes nonrenormalizable).  The finiteness due to the vanishing of the pole part of a formally divergent contribution because of its chirality, with the surviving of the nonpolynomial, logarithmic-like contributions, has never been achieved in superfield theories formulated in terms of ${\cal N}=1$ superfields.  It is easy to check that this contribution to the effective action is scale invariant. Earlier, the arising of the logarithm-like finite contributions to the effective action with the absence of divergences has been shown to take place only in  theories formulated in ${\cal N}=2$ superspace, more precisely, in ${\cal N}=4$ super-Yang-Mills theories \cite{n4sym}. So, we showed a new mechanism for achieving the finiteness for the ${\cal N}=1$ superfield theories.
We also saw that  in a certain limit the one-loop effective potential we found reproduces the result for the Wess-Zumino model. 

{\bf Acknowledgements.} This work was partially supported by Conselho
Nacional de Desenvolvimento Cient\'{\i}fico e Tecnol\'{o}gico (CNPq)
and Funda\c{c}\~{a}o de Amparo \`{a} Pesquisa de Estado de S\~{a}o
Paulo  (FAPESP).
A. Yu. P. has been supported by the CNPq project 303438-2012/6. F. S. Gama has been supported by the CNPq process No. 141228/2011-3.

\end{document}